\documentclass[reprint,prd]{revtex4-2}
\usepackage{hyperref}
\hypersetup{colorlinks,allcolors=black}
 \usepackage{graphicx}
 \usepackage{float}
 \usepackage{amsmath}
 \usepackage{amssymb}
 \usepackage[utf8]{inputenc}
 \usepackage{natbib}
 \usepackage{slashed}

\begin{document}
\title{Restoration of the Lorentz symmetry of particle propagator in the ghost condensate model}
\author{Suppanat Supanyo$^1$}
\author{Pitayuth Wongjun$^1$}
\author{Lunchakorn Tannukij$^2$}
\affiliation{$^1$The Institute for Fundamental Study (IF), Naresuan University, Phitsanulok 65000, Thailand}
 \affiliation{$^2$Department of Physics, School of Science, King Mongkut's Institute of Technology Ladkrabang, Bangkok 10520, Thailand}
\begin{abstract}
    The ghost condensate model has been proposed to protect the vacuum state of the cosmological phantom model yielding the gradient instability regime and the spontaneous Lorentz symmetry breaking in the quantum level with the dispersion relation $\omega^2\sim k^4$. In this paper, we point out that the Lorentz symmetry in the  propagator can be restored if the explicit Lorentz symmetry-breaking process is promoted. The particle excitation of the phantom field can propagate through spacetime with the Lorentzian dispersion relation $\omega^2=k^2$ in arbitrary background value. Moreover, the tree-level gradient instability regime can be removed under the characteristic Lorentz violation source of ultraviolet physics.
\end{abstract}
\maketitle

\section{Introduction}
From observations of the accelerated expansion of the universe, the phantom scalar field \cite{CALDWELL200223, Ludwick:2017tox,PhysRevD.68.023509, PhysRevD.71.023515,PhysRevD.72.023003, Nojiri:2005pu,Nojiri:2006ww,e23040404} has been proposed as one of the viable models to account for dark energy in the regime where the equation of state parameter satisfies $w<-1$. However, this scenario is not without theoretical challenges. Although the negative kinetic energy associated with the phantom scalar field can be phenomenologically useful in cosmology, it also introduces severe issues regarding the stability of the model at both the classical and quantum levels. 

To address these problems, a wide variety of frameworks have been proposed under the assumption of Lorentz symmetry, motivated by different theoretical perspectives. For instance, from a mathematical standpoint or within string-inspired constructions, models such as k-essence \cite{PhysRevD.103.043518,PhysRevD.63.103510,ARMENDARIZPICON1999209} and Dirac-Born-Infeld (DBI) action \cite{PhysRevD.83.101301,PhysRevD.70.123505,PhysRevD.77.023511} introduce non-canonical kinetic terms for the scalar field. These modifications allow the models to better accommodate cosmological observations while improving classical stability. Additionally, the quintom model \cite{PhysRevD.71.047301,PhysRevD.73.123509,Cai:2009zp,Yang:2024kdo} introduces two dynamical degrees of freedom—one resembling quintessence and the other phantom to alleviate the inherent instabilities of the pure phantom scenario. Furthermore, within the framework of effective field theory, the ghost condensate model (GC) \cite{Arkani-Hamed:2003pdi,Arkani-Hamed:2003juy, Anisimov:2004sp,Piazza:2004df,Arkani-Hamed:2005teg, Bhattacharya:2014pta}  offers an alternative mechanism to stabilize classical and quantum instabilities by incorporating the Lorentz invariant higher dimension operators into the effective Lagrangian. However, this comes at the cost of spontaneous Lorentz symmetry breaking in the low-energy  quantum level, while the ultraviolet (UV) completion is typically assumed to remain Lorentz invariant. 

Recent experimental searches for extensions of the Standard Model \cite{Colladay:1996iz,Colladay:1998fq} and precision tests of Lorentz invariance 
have so far revealed no conclusive evidence for any violation of Lorentz symmetry \cite{D0:2012rbu,Carle:2019ouy,CMS:2024rcv}. However, at the earliest stages of the universe, where individual particles could reach extremely high energies, it remains an open question whether Lorentz symmetry is indeed a fundamental property of nature. It is conceivable that Lorentz invariance may not exist at the ultraviolet (UV) scale, but could \textit{emerge approximately} in the infrared (IR) regime, where effective field theories exhibit approximate Lorentz symmetry even if it is absent at the fundamental level.

In this work, under the above hypothesis, we develop an improved formulation of the ghost condensate model. We demonstrate that if Lorentz symmetry is not a fundamental property of the ultraviolet (UV) completion of the phantom scalar field theory, it is possible to construct a phantom model that achieves improved stability at both the classical and quantum levels. Moreover, in such a framework, an effective restoration of Lorentz invariance can emerge in the low-energy particle excitations of the phantom field, even if the underlying UV theory explicitly breaks Lorentz symmetry.

This paper is organized as follows. 
In Sec.~II, we review the mechanism of Lorentz symmetry breaking 
within the ghost condensate (GC) framework. 
In Sec.~III, we introduce our modified GC model, 
in which explicit Lorentz symmetry breaking is incorporated 
at the ultraviolet (UV) scale. 
In Sec.~IV, we analyze the one-loop corrections to the phantom propagator, 
examine the implications for UV symmetries, 
and discuss several open problems. 
Finally, our conclusions are summarized in Sec.~V

\section{Review on the Lorentz symmetry breaking of Ghost condensate in flat spacetime}
In this section,  the GC model is reviewed by considering the toy Lagrangian in the flat Minkowski spacetime, 
\begin{align}\label{7}
    \mathcal{L}_{GC}=-\frac{1}{2}\partial_\mu\phi\partial^\mu\phi+\frac{1}{4M^4}\left(\partial_\mu\phi\partial^\mu\phi\right)^2,
\end{align}
instead of the generic function $\mathcal{L}_{GC}=P(\partial_\mu\phi\partial^\mu\phi)$, to simplify the analytic calculation.

\begin{figure}[h]
    \centering
    \includegraphics[width=1\linewidth]{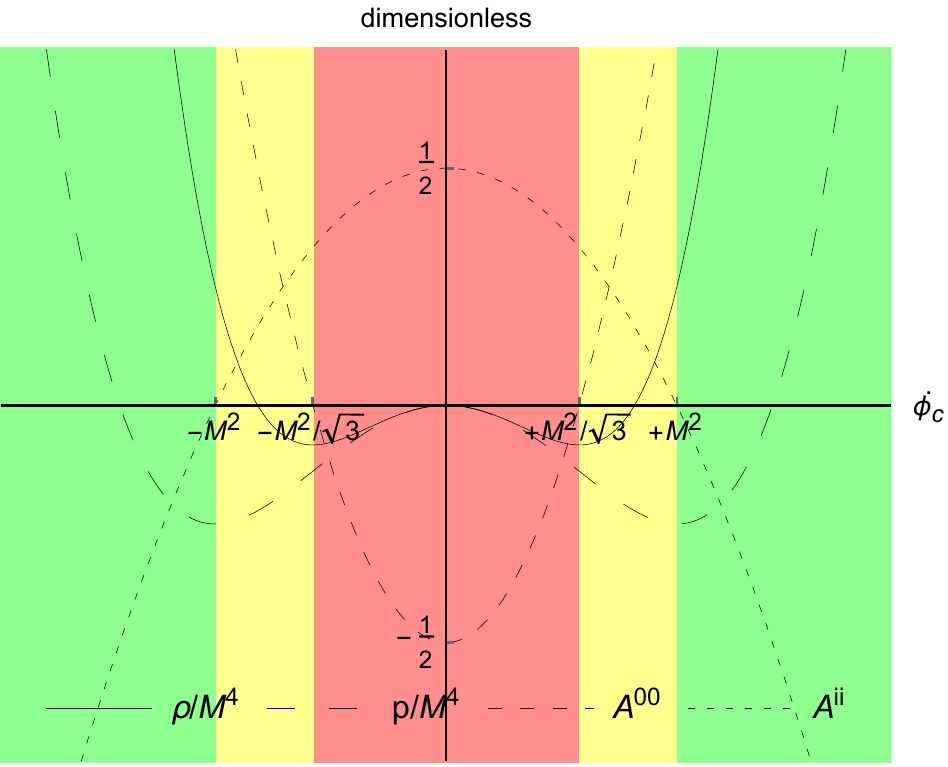}
    \caption{The energy density, the pressure, the coefficients of the ghost condensate model in a function of the constant $\dot{\phi}_c$}
    \label{fig:fig3}
\end{figure}

By assuming an isotropic background and homogeneous, the phantom field can be separated as  
\begin{align}\label{GC1}
\phi(t,\boldsymbol{x})=\phi_b(t)+\pi(t,\boldsymbol{x})
\end{align}
where $\pi$ is the perturbed field (quantized field, or the field of particle excitation) including the spatial fluctuation and $\phi_b$ is the classical background field. Substituting Eq.~\eqref{GC1} into Eq.~\eqref{7}, we can obtain the equation of motion (EOM), the energy density and pressure of the phantom field, respectively,
\begin{align}
  &\frac{\partial}{\partial t}\left[ \left(1-\frac{\dot{\phi}_b^2}{M^4}\right)\dot{\phi}_b\right]=\left(1-\frac{3\dot{\phi}_b^2}{M^4}\right)\ddot{\phi}_b=0,\label{GCEOM}
\\
    &\rho_{GC}=-\frac{\dot{\phi}_b^2}{2}+\frac{3\dot{\phi}_b^4}{4M^4},\label{rhoGC}
    \\
    & p_{GC}=-\frac{\dot{\phi}_b^2}{2}+\frac{\dot{\phi}_b^4}{4M^4}.\label{pGC}
\end{align}
From Eq.~\eqref{GCEOM}, $\dot{\phi}_b(t)=constant=\dot{\phi}_c$ is a solution leading to 
 \begin{align}\label{GCb}
     \phi_b(t)=\phi_c+\dot{\phi}_c t.
 \end{align}
Here, $\rho_{GC}$ and $p_{GC}$ are bounded from below with the minimum points $\dot{\phi}_c=M^2/\sqrt{3}$ and $\dot{\phi}_c=M^2$, respectively. Both minimum points are also the solutions of Eq.~\eqref{GCEOM}. At $\dot{\phi}_c=M^2/\sqrt{3}$, it is a stationary point but this point leads to a negative energy contribution. The original work of GC expects that the phantom field in the Friedmann–Lemaître–Robertson–Walker metric evolves into the saddle point of Lagrangian, which relates to the minimum point of pressure at $\dot{\phi}_c=M^2$. This stationary point provides the positive energy state.
Then, using Eqs.~\eqref{GCb} and \eqref{GC1}, the Lagrangian \eqref{7} is reorganized as 
\begin{align}
\mathcal{L}_{GC}=&-\frac{\dot{\phi}_c^2}{2}+\frac{\dot{\phi}_c^4}{4M^4}+V_0+A^{\mu\nu}\partial_\mu\pi\partial_\nu\pi-Y\dot{\pi}\nonumber
\\
&+\mathcal{L}_{GC}^{(3\pi)}+\mathcal{L}_{GC}^{(4\pi)}
\end{align}
where $A^{0i}=A^{i0}=A^{ij}=0$,
\begin{align}
    A^{00}=-\frac{1}{2}+\frac{3\dot{\phi}_c^2}{2M^4},
     A^{ii}=+\frac{1}{2}-\frac{\dot{\phi}_c^2}{2M^4},
     Y=\dot{\phi}_c-\frac{\dot{\phi}_c^3}{M^4}.
\end{align}
The $\dot{\pi}$ term represents the vacuum decay term in the quantum field theory. This term can vanish in an arbitrary background without a specific value of $\dot{\phi}_c$. By reorganizing
\begin{align}
    Y\dot{\pi}=Y a^\mu \partial_\mu\pi,
\end{align}
where $a^{\mu}=(1,0,0,0)$, this term is obviously a boundary term that vanished in the action level by the integration by part
\begin{align}
    \int d^4x Y a^\mu \partial_\mu\pi= \int d^4x \partial_\nu(Ya^{\nu}\pi)=0.
\end{align}
Hence, the value of $\phi_c$ can be an arbitrary value and give rise to the different quantum field shown in Fig.~\ref{fig:fig3}.
In our analysis, the word ``kinetic term'' refers to $A^{\mu\nu}\partial_\mu\pi\partial_\nu\pi$, the word ``'kinetic energy term' refers to $\dot{\pi}^2$, and the word ``gradient kinetic energy term'' refers to $(\nabla\pi)^2$.

\begin{itemize}
    \item $0<|\dot{\phi}_c|\leq M^2/\sqrt{3}$: We have  $|A^{00}|\neq|A^{ii}|$, $-1/2<A^{00}\leq 0$ and $+1/2<A^{ii}\leq+1/3$ so Lorentz symmetry is broken due to the non-covariant form of the kinetic term. According to the negative sign of $A^{00}$, the excitation of field $\pi$ is a ghost particle with dispersion relation $\omega^2=constant\times \boldsymbol{k}^2$.

    \item $M^2/\sqrt{3}<|\dot{\phi}_c|<M^2$: We have  $|A^{00}|\neq|A^{ii}|$, $0<A^{00}<+1$, $+1/3>A^{ii}>0$. The ghost particle $\pi$ condenses into the particle with positive kinetic energy. However, this case, so-called the gradient instability, provides the classical exponential growth solution, leading to an unhealthy quantum field theory.
    The merit of this regime is that it potentially provides the equation of state (EOS) $w<-1$ \cite{Hsu:2004vr}. 

    \item $|\dot{\phi}_c|\geq M^2$: In this regime, $A^{00}\geq+1$ and $A^{ii}\leq 0$,  the quantum theory is healthy. The exciting field $\pi$ is the particle with positive kinetic energy. The gradient kinetic energy receives a correct sign.  $\pi$ can propagate in the spacetime with the dispersion relation $\omega^2-constant\times \boldsymbol{k}^2$=0. At $\dot{\phi}_c=M^2$, this background value is a stationary point of the Lagrangian.  However, at this point, the gradient kinetic energy term $(\nabla\pi)^2$ vanish due to $A^{ii}=0$. This theoretical problem can be fixed by rewriting the Lagrangian in a more general way. The effective operator $(\Box^2\phi)^2$ can be added into the Lagrangian \eqref{7} while the background solution \eqref{GCb} is still settles. Then, the gradient kinetic energy is recovered as
\begin{align}
    \mathcal{L}_{GC}\supset\dot{\pi}^2-\frac{(\nabla^2\pi)^2}{M^4}
\end{align}
Here,  $\pi$ has a positive kinetic energy particle with the non-Lorentzian dispersion relation $\omega^2\sim k^4/M^2$.
\end{itemize}

In short, the positive energy condition and vacuum stability of the phantom field model can be protected by the ghost condensate model, which requires the new physics respecting the Lorentz symmetry at the energy scale $M$, see the Lorentz invariant higher dimension operator in Eq.~\eqref{7}.  All regimes of $\dot{\phi}_b$(but $\dot{\phi}_b=0$) lead to the Lorentz symmetry breaking due to $A^{00}\neq-A^{ii}$.

\section{Explicit Lorentz symmetry breaking in the ghost condensate model from the Lorentz violation operator}
Let us assume that the phantom field couples with a Lorentz violation (LV) source from UV theory, which becomes relevant at the energy scale $M_{LV}$.
 One can assume that the effective Lagrangian for the phantom field can be written as
\begin{align}\label{1}
    \mathcal{L}=&-\frac{1}{2}\eta^{\mu\nu}\partial_\mu\phi\partial_\nu\phi+\frac{1}{4M_{LV}^4}\left(B^{\mu\nu}\partial_\mu\phi\partial_\nu\phi\right)^2,
\end{align}
where $M_{LV}$ is the cutoff scale of this effective theory and $B^{\mu\nu}$ is the LV tensor defined as 
\begin{align}
    B^{\mu\nu}=\begin{pmatrix}
       B^{00} & 0 & 0 & 0 \\
       0 & B^{11} & 0 & 0 \\
       0 & 0 & B^{22} & 0 \\
       0 & 0 & 0 & B^{33}
    \end{pmatrix}.
\end{align}
For $B^{00}=-B^{ii}=1$, this model is returned to the standard ghost condensate model.
Under the condition $B^{00}\neq B^{11}\neq B^{22}\neq B^{33}$, it is easy to see that this Lagrangian \eqref{1} is not invariant under the Lorentz transformation $x'^\mu=\Lambda^\mu{}_\nu x^{\nu}$. The quadratic part is in the covariant form which is Lorentz invariant quantity. On the other hand, the quartic part of Lagrangian breaks the Lorentz symmetry, explicitly. In the small momentum limit $(\partial\phi)^4\ll M_{LV}^4$, the Lorentz symmetry of the phantom field is restored.  

\begin{figure}[h]
    \centering
    \includegraphics[width=1\linewidth]{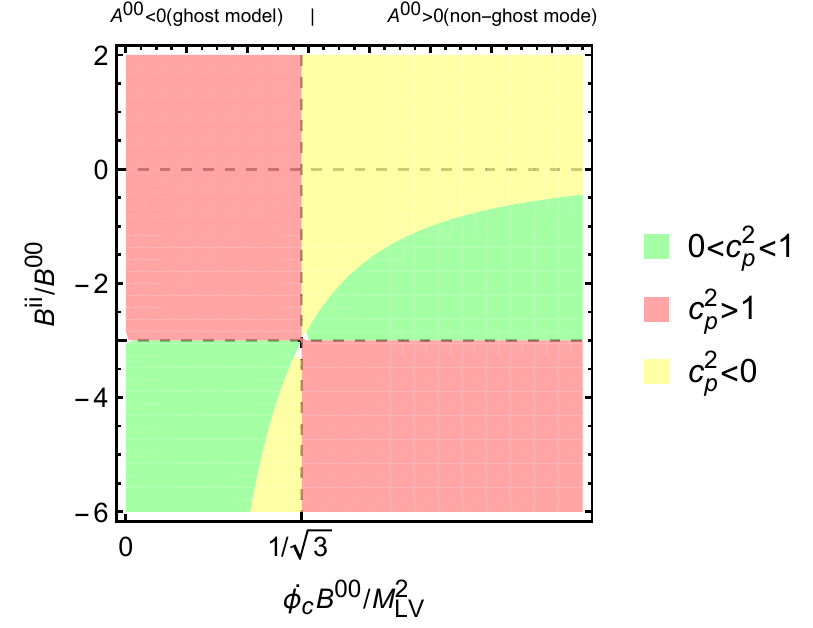}
    \caption{The constraint on $c_p^2$ for arbitrary value of $B^{ii}/B^{00}$ and $\dot{\phi}_c$.}
    \label{fig:fig7}
\end{figure}

\begin{figure}[h]
    \centering
    \includegraphics[width=1\linewidth]{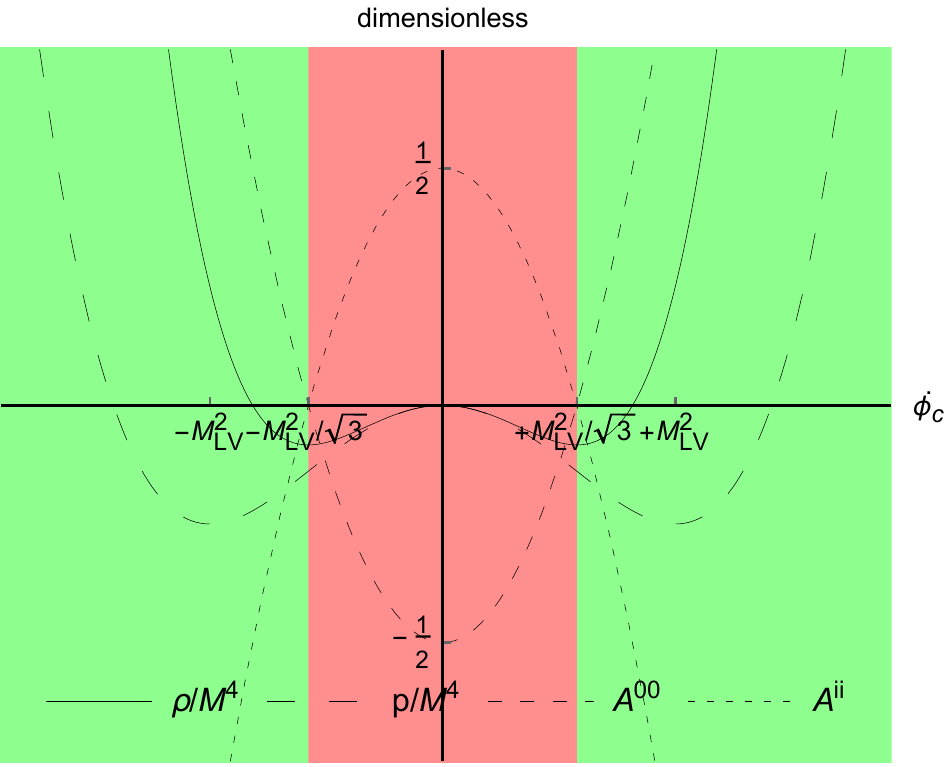}
    \caption{The sketch of the energy density, the pressure, the coefficients of the explicit Lorentz symmetry breaking ghost condensate model in a function of the constant $\dot{\phi}_c$ with $B^{00}=1$}
    \label{fig:fig4}
\end{figure}

In this section, we are going to show that, under the condition   
\begin{align}
    B^{00}\neq -B^{ii},
\end{align}
it is possible to construct the new ghost condensate model that the particle excitation can propagate with the dispersion relation $\omega^2=\boldsymbol{k}^2$ for arbitrary background value $\dot{\phi}_c$. Moreover, the gradient instability regime can be removed.  To simplify this model, we also add the isotropic property yielding 
\begin{align}
   B^{11}=B^{22}=B^{33}.
\end{align}
Then,  the phantom field can be separated into the classical homogeneous background $\phi$ and the excitation of field $\chi$ including the spatial fluctuation as
\begin{align}\label{sepfield}
    \phi(t,\boldsymbol{x})=\phi_b(t)+\chi(t,\boldsymbol{x}).
\end{align}
The EOM of the classical homogeneous background is
\begin{align}
    \left(1-\frac{3\dot{\phi}_b^2}{M_{LV}^4}\right)\ddot{\phi}_b=0,
\end{align}
leading to the same classical solution as
\begin{align}\label{solutionELGC}
    \phi_b=\phi_c+\dot{\phi}_c t,
\end{align}
where $\phi_c$ and $\dot{\phi}_c$ are constant. Using Eqs.~\eqref{sepfield} and \eqref{solutionELGC}, the energy density and pressure of \eqref{1} in the homogeneous background are given by
\begin{align}
    \rho=-\frac{\dot{\phi}_c^2}{2}+\frac{3(B^{00})^2\dot{\phi}_c^4}{4M_{LV}^4},
    \\
     p=-\frac{\dot{\phi}_c^2}{2}+\frac{(B^{00})^2\dot{\phi}_c^4}{ 4M_{LV}^4},
\end{align}
and the Lagrangian \eqref{1} can be reorganized as
\begin{align}\label{lchiEGC}
    \mathcal{L}=&-\frac{\dot{\phi}_c^2}{2}+\frac{\dot{\phi}_c^2}{4M_{LV}^4}+A^{\mu\nu}\partial_\mu\chi\partial^\mu\chi-Y\dot{\chi}\nonumber
    \\
    &+\mathcal{L}^{(3\chi)}+\mathcal{L}^{(4\chi)}
\end{align}
where 
\begin{align}
   A^{00}=&-\frac{1}{2}+\frac{3(B^{00})^2\dot{\phi}_c^2}{2M_{LV}^4},\label{A00EGC}
    \\
     A^{ii}=&+\frac{1}{2}+\frac{B^{00}B^{ii}\dot{\phi}_c^2}{2M_{LV}^4},\label{AiiEGC}
     \\
     Y=&\dot{\phi}_c-\frac{(B^{00})^2\dot{\phi}_c^3}{M_{LV}^4}.\label{YEGC}
\end{align}
The $Y$ term can be vanished by integration by part as shown in the previous section.
The kinetic term of $\chi$ depends on the choice of the Lorentz violation tensor $B^{\mu\nu}$. From Eqs.~\eqref{A00EGC}-\eqref{AiiEGC}, there are two different mode of $\chi$ depending on $\dot{\phi}_c$ as follows.
\begin{itemize}
    \item $|\dot{\phi}_c|<M_{LV}^2/\sqrt{3}B^{00}$:  The field $\chi$ is ghost particle since $A^{00}<0$.

    \item $|\dot{\phi}_c|>M_{LV}^2/\sqrt{3}B^{00}$:   The field $\chi$ is not ghost particle since $A^{00}>0$.
\end{itemize} 
In both cases, the particle excitation propagates with the dispersion relation
\begin{align}
    \omega^2=-\frac{A^{ii}}{A^{00}}\boldsymbol{k}^2.
\end{align}
With this dispersion relation, we can define the speed of the particle propagation by the group velocity
\begin{align}
    c_p^2=\left(\frac{\partial\omega}{\partial\boldsymbol{k}}\right)^2=-\frac{A^{ii}}{A^{00}},
\end{align}
depending on $B^{00}$, $B^{ii}$, and $\dot{\phi}_c$ illustrated in Fig.~\ref{fig:fig7}.
The propagation speed of $\chi$ can be separated into three regimes: 1. superluminal speed (red shading regime), 2. below the speed of light (green shading regime), and 3. instability (yellow shading regime). The analysis is shown below.
\begin{itemize}
    \item $B^{ii}/B^{00}>0$: In the range $|\dot{\phi}_c|<M_{LV}^2/\sqrt{3}B^{00}$,  $\chi$ is a ghost particle with the superluminal speed, so the causality is violated. In the range $|\dot{\phi}_c|>M_{LV}^2/\sqrt{3}B^{00}$, $\chi$ has a correct sign kinetic energy, however, the gradient kinetic energy obtains a wrong sign leading to the gradient instability problem.

    \item $-3<B^{ii}/B^{00}<0$: In the range $|\dot{\phi}_c|<M_{LV}^2/\sqrt{3}B^{00}$,  $\chi$ is ghost particle propagating with the superluminal speed.  In the range $|\dot{\phi}_c|>M_{LV}^2/\sqrt{3}B^{00}$,  $\chi$ can obtain positive kinetic energy with $c_p^2<1$. Therefore, we can obtain a theory obeying causality in the positive kinetic energy regime but breaking causality in the negative kinetic energy regime.

    \item $B^{ii}/B^{00}<-3$:   We find that, in the range $|\dot{\phi}_c|<M_{LV}^2/\sqrt{3}B^{00}$,  the  $\chi$ is ghost particle propagating below the speed of light. On the other hand, in the range $|\dot{\phi}_c|>M_{LV}^2/\sqrt{3}B^{00}$, $\chi$ obtains positive kinetic energy propagating superluminally in spacetime. In this regime, we can obtain a theory obeying causality in the negative kinetic energy regime but breaking causality in the positive kinetic energy regime.
\end{itemize}
Among various possible phenomena depending on the ratio $B^{ii}/B^{00}$, we find that there is some special value of the Lorentz violation tensor $B^{\mu\nu}$ to construct a healthy ghost condensate theory. To restore the Lorentz symmetry in the quadratic part, the kinetic energy term has to be reorganized in the form 
\begin{align}
A^{\mu\nu}\partial_\mu\chi\partial^\mu\chi=\text{constant}\times (\dot{\chi}^2-(\nabla\chi)^2)
\end{align}
Therefore, the condition  
\begin{align}
  A^{00}=-A^{ii}
\end{align}
is desired. As a consequence, the tensor $B^{\mu\nu}$ must have a characteristic property
\begin{align}
    B^{ii}=-3B^{00},
\end{align}
yielding
\begin{align}\label{b3}
    B^{\mu\nu}=B^{00}\begin{pmatrix}
      1 & 0 & 0 & 0 \\
       0 & -3 & 0 & 0 \\
       0 & 0 & -3 & 0 \\
       0 & 0 & 0 & -3
    \end{pmatrix}.
\end{align}
With this characteristic choice of the Lorentz violation tensor,  the quadratic part of $\chi$ always preserves the Lorentz symmetry independent with the classical background $\phi_c$, 
\begin{align}
    A^{00}=-A^{ii}=-\frac{1}{2}+\frac{3B^{00}\dot{\phi}_c^2}{2M_{LV}^4},
\end{align}
clearly illustrated in Fig.~\ref{fig:fig4}.
\noindent Here,  the gradient instability and the Lorentz violation in the kinetic term of $\chi$ are completely removed. $\chi$  propagates with the dispersion relation
\begin{align}
    \omega^2=-\frac{A^{ii}}{A^{00}}\boldsymbol{k}^2=\boldsymbol{k}^2,
\end{align}
for all values of the background $\dot{\phi}_c$. An unhealthy situation such as the non-vanishing of the gradient kinetic energy at the stationary point of Lagrangian in the traditional GC model does not exist, thus the term such as $(\Box\phi)^2$ is not necessary to fix the tree level propagator of $\chi$. 

Even though the Lorentz violation is eliminated from the propagator, it is still explicitly illustrated in the self-interaction terms as
\begin{align}
&\mathcal{L}^{(3\chi)}=\frac{B^{00}\dot{\phi}_c}{M^4_{LV}}(a^{\alpha}\partial_{\alpha}\chi )(B^{\mu\nu}\partial_\mu\chi\partial_\nu\chi)\label{lc3}
    \\
   & \mathcal{L}^{(4\chi)}=\frac{1}{4M^4_{LV}}(B^{\mu\nu}\partial_\mu\chi\partial_\nu\chi)^2\label{lc4}
\end{align}
where $a^{\nu}=(1,0,0,0)$.
 In the limit $\omega^2\ll M_{LV}^2$, the self-interaction terms become irrelevant in the perturbative calculation so the Lagrangian remains the Lorentz invariant kinetic energy of $\chi$ 
 \begin{align}
    \mathcal{L}\simeq&-\frac{\dot{\phi}_c^2}{2}+\frac{\dot{\phi}_c^4}{4M_{LV}^4}+A^{00}\eta^{\mu\nu}\partial_\mu\chi\partial^\mu\chi,
\end{align}
 where $\dot{\phi}_c$ is just a constant. Therefore, the Lorentz symmetry is approximately restored in the low-energy regime.  The field $\chi$ can be fluctuated between ghost and non-ghost depending on $\dot{\phi}_c$.

\section{Discussion}
\subsection{Counterterm from one-loop contribution}
We consider the essential counterterm from the one-loop correction contribution to the $\chi$ propagator.
 From the interactions in Eqs.~\eqref{lc3}-\eqref{lc4}, the Feynman rules are
\begin{align}
&iG_{tree}=\frac{i}{A^{00}\eta_{\mu\nu} p^{\mu}p^{\nu}},
\\
&V_{3\chi\to 0}(p_1,p_2,p_3)=\nonumber
\\
&-\frac{2B^{00}\dot{\phi}_c}{M^4_{LV}}a_{\alpha}B_{\mu\nu} \left(p_1^{\alpha}p_2^{\mu}p_3^{\nu}+p_2^{\alpha}p_1^{\mu}p_3^{\nu}+p_3^{\alpha}p_1^{\mu}p_2^{\nu}\right),
\\
&V_{4\chi\to 0}(p_1,p_2,p_3,p_4)=\nonumber
\\
&\frac{2i}{M^4_{LV}}B_{\mu\nu}B_{\alpha\beta}\left(p_1^\mu p_2^{\nu}p_3^\alpha p_4^\beta+p_1^\mu p_3^{\nu}p_2^\alpha p_4^\beta+p_1^\mu p_4^{\nu}p_2^\alpha p_3^\beta\right).
\end{align}
where $B_{\mu\nu}=\eta_{\mu\alpha}\eta_{\nu\beta}B^{\alpha\beta}$ and $a_{\mu}=\eta_{\mu\nu}a^{\nu}$.
The propagator from the sum overall 1 particle irreducible diagram is
\begin{align}
    iG=\frac{i}{A^{00}p^2-\Pi(p^2)}.
\end{align}
 $\Pi(p^2)$ can be separated in terms of one loop self-energy amplitude from the three-legged vertex and four-legged vertex as
 \begin{align}
     i\Pi(p^2)=i\Pi_3(p^2)+i\Pi_4(p^2).
 \end{align}
 where
 \begin{align}
    &i\Pi_3(p^2)=\nonumber
    \\
    &\int \frac{d^4l}{(2\pi)^4}\frac{i^2V_{3\chi\to0}(l,p,-l-p)V_{3\chi\to0}(-l,-p,l+p)}{(A^{00})^2l^2(l-p)^2} ,
    \\
     &i\Pi_4(p^2)=\int \frac{d^4l}{(2\pi)^4}\frac{iV_{4\chi\to 0}(l,-l,p,-p)}{A^{00}l^2}.
 \end{align}
 By evaluating this loop integral with the dimensional regularization ($d^4l\to d^{4-2\varepsilon}l$), $i\Pi_4$ traditionally vanishes in the massless limit while $i\Pi_3$ remains non-zero. The UV-divergence term of $i\Pi(p^2)$ is expressed as
 \begin{align}
     \left[i\Pi(p^2)\right]_{UV}= \frac{i}{960\pi^2 (A^{00})^2M_{LV}^8\varepsilon}f(\omega,\boldsymbol{k}),
 \end{align}
 where
 \begin{align}
     f(\omega,\boldsymbol{k})=36(B^{00})^4\dot{\phi}_c^2\omega^2(43 \boldsymbol{k}^4-50\boldsymbol{k}^2\omega^2+15\omega^4).
 \end{align}
 To absorb the UV-divergence proportional to (energy)$^6$ contributing to $\chi$ propagator, the term $BBBB\dot{\phi}_c^2 \partial\partial\partial\partial\chi\partial\partial\chi/M_{LV}^8$ is necessary. From the power counting formula \cite{Manohar:1983md,Gavela:2016bzc}, this term is dimension-12 operator that should contain $\phi^4$ and 8-$\partial_\mu$. Hence, the counterterm should be given in the form 
 \begin{align}
     \mathcal{L}_{ct}=&\frac{c}{M_{LV}^8} B^{\mu\nu}B^{\alpha\beta}B^{\kappa\sigma}B^{\delta\rho}\partial_{\mu}\partial_{\nu}\phi \partial_{\alpha}\partial_{\beta}\partial_{\kappa}\partial_{\sigma}\phi \partial_\delta\phi\partial_\rho\phi\nonumber
     \\
     &+\text{permutation}~.
 \end{align}
Under the classical solution $\phi=\dot{\phi}_ct$, the counterterm Lagrangian vanishes, so the one-loop counterterm does not contribute to the classical energy and pressure function of the theory.  Therefore, the classical stability is protected against the 1-loop contribution. After removing the divergence, the self-energy function is 
\begin{align}
    i\Pi= i \frac{ f(\omega,\boldsymbol{k}) g(p^2)}{
   M_{\text{LV}}^8},
\end{align}
where 
\begin{align}
  g(Q^2)  =\frac{\left(15 \log \left(\frac{\mu ^2}{Q^{2}}\right)+46\right)}{14400 \pi ^2}
\end{align}
and $Q^2=-p^2$. The propagator of $\chi$ is given by
\begin{align}
    iG=\frac{i}{\left(2A^{00}+\frac{f(\omega,\boldsymbol{k})}{M_{LV}^8}\right)\omega^2-\left(2A^{00}+n\frac{f(\omega,\boldsymbol{k})}{M_{LV}^8}\right)\boldsymbol{k}^2},
\end{align}
where $n=-(25+\sqrt{1270})/15\simeq-4$. The coefficients in front of $\omega^2$ and $\boldsymbol{k}^2$ are slightly different due to self-energy correction.
In the limit $\boldsymbol{k}^2\ll M^2_{LV}$ and $\omega^2\ll M^2_{LV}$, the Lorentz symmetry of this propagator is an approximated restoration. However, the Lorentz
violation of the propagator can be slightly induced from
the loop level. In the limit  $\boldsymbol{k}^2\simeq\omega^2\simeq M_{LV}^2$,  the 1-loop effect  becomes 
strongly relevant at the explicit symmetry breaking scale. 

\subsection{Characteristic source of the Lorentz violation}
In this model, the Lorentz symmetry in the propagator of $\chi$ is restored when the UV physics contributing to the GC model is embedded with characteristic tensor $B^{\mu\nu}$ in Eq.~\eqref{b3}. The effective interaction term
\begin{align}\label{lvin}
\frac{(B^{\mu\nu}\partial_\mu\phi\partial_\nu\phi)^2}{4M_{LV}^4}
\end{align}
is invariant under the modified Lorentz transformation with shifting 
\begin{align}
    c\to 3c
\end{align}
for instance, in $x$-direction,
\begin{align}
    t'=\frac{t-v\frac{x}{(3c)^2}}{\sqrt{1-\left(\frac{v}{3c}\right)^2}},~
    x'=\frac{x-vt}{\sqrt{1-\left(\frac{v}{3c}\right)^2}},~y'=y,~z'=z.
\end{align}
Under this new symmetry,  the characteristic property of this UV theory allows the speed of the signal transition to be greater than the speed of light, which exists at the interaction level.  However, with the current experimental observation at the Large Hadron Collider, the sign of the Lorentz violation has never been reported in the TeV \cite{D0:2012rbu,Carle:2019ouy,CMS:2024rcv}. The particle $\chi$ can possibly couple with the SM particle above the TeV scale, for example, inducing via a gravitation exchange, discussed in \cite{Arkani-Hamed:2003juy}. In this reference, the higher-dimension operator from integrating out graviton degree of freedom is suppressed by the Planck mass $\hat{O}_n(\chi,\text{SM particles})/M_P^{n-4}$. The Lorentz violation signal can appear in an arbitrarily high-energy regime, while the scale of parameter $M_{LV}$ can be arbitrarily high or arbitrarily low into the cosmological constant scale. If $\chi$ couples with the SM solely via gravitational interaction, consequently the model can survive. One might seek to investigate the cosmological phenomena, which lie beyond the scope of this study. Nevertheless,  before concluding this paper, we highlight open problems in gravitational interactions.

\subsection{Open problem on \textit{the Gravity coupling}}
In general, the action of field theory in the flat spacetime can be promoted into the curved spacetime by $\int d^4 x\to\int d^4x\sqrt{-g(x)}$, $\partial_\mu\to\nabla_\mu$, and
\begin{align}
  \eta^{\mu\nu}  \to g^{\mu\nu}(x),~\eta_{\mu\nu}  \to g_{\mu\nu}(x).
\end{align}
One of our questions is how to deal with the self-interaction term, including the Lorentz violation tensor
\begin{align}
    \frac{(B^{\mu\nu}\partial_\mu\phi\partial_\nu\phi)^2}{4M_{LV}^4}
\end{align}
in the Lagrangian \eqref{1}. This term can be parametrized in terms of $\eta$ in several independent ways, for example,
\begin{align}
    B^{\mu\nu}\partial_\mu\phi\partial_\nu\phi=&(\eta^{\mu\nu}+b^{\mu\nu})\partial_\mu\phi\partial_\nu\phi,
    \\
B^{\mu\nu}\partial_\mu\phi\partial_\nu\phi=&\eta^{\mu\alpha}\eta^{\nu\beta}B_{\alpha\beta}\partial_\mu\phi\partial_\nu\phi,
\end{align}
where diag$(b^{\mu\nu})=(-1+B^{00},1-3B^{00},1-3B^{00},1-3B^{00})$. 
The last one can think that the tensor $B$ is not independent of the metric tensor $\eta$
\begin{align}
    B^{\mu\nu}\not\subset\eta^{\mu\nu},
\end{align}
which means that the tensor $B$ will not be promoted in terms of $g$ in the curved spacetime. Every example case can lead to different gravity-GC actions,
\begin{align}
    &S_1=\int d^4x\sqrt{-g}\nonumber
    \\
    &\times\left(-\frac{1}{2}g^{\mu\nu} \partial_\mu\phi\partial_\nu\phi+\frac{((g^{\mu\nu}+b^{\mu\nu})\partial_\mu\phi\partial_\nu\phi)^2}{4M_{LV}^4}\right),\label{s1}
    \\
    & S_2=\int d^4x\sqrt{-g}\nonumber
    \\
     &\times\left(-\frac{1}{2}g^{\mu\nu} \partial_\mu\phi\partial_\nu\phi+\frac{(g^{\mu\alpha}g^{\nu\beta}B_{\alpha\beta}\partial_\mu\phi\partial_\nu\phi)^2}{4M_{LV}^4}\right), \label{s2}
     \\
      &S_3=\int d^4x\sqrt{-g}\nonumber
      \\
     &\times  \left(-\frac{1}{2}g^{\mu\nu} \partial_\mu\phi\partial_\nu\phi+\frac{(B^{\mu\nu}\partial_\mu\phi\partial_\nu\phi)^2}{4M_{LV}^4}\right), \label{s3}
\end{align}
yielding a different energy density 
\begin{align}
    \rho_1=&-\frac{\dot{\phi}_b^2}{2}+\frac{B^{00}(2-B^{00})\dot{\phi}_b^4}{4M_{LV}^4},
    \\
    \rho_2=&-\frac{\dot{\phi}_b^2}{2}+\frac{3(B^{00})^2\dot{\phi}_b^4}{4M_{LV}^4},
    \\
    \rho_3=&-\frac{\dot{\phi}_b^2}{2}-\frac{(B^{00})^2\dot{\phi}_b^4}{4M_{LV}^4},
\end{align}
evaluated by $T_{00}=2\partial \mathcal{L}/\partial g^{00}-g_{00}\mathcal{L}$, but the pressure is obtained in the same form
\begin{align}
    p=-\frac{\dot{\phi}_b^2}{2}+\frac{(B^{00})^2\dot{\phi}_b^4}{4M_{LV}^4}.
\end{align}
Obviously, there could be three distinct equations of state parameters $w=p/\rho$, providing different theoretical predictions of the accelerating expansion of the universe. These predictions depend on how the UV physics, embedded with LV, couples to the phantom scalar field. Another important discussion is the conservation of energy-momentum.  For those three distinct LV cases, the energy-momentum conservation is no longer applicable. This is obviously due to the explicit Lorentz violation in the action. Interestingly, the non-conservation part in three cases can be characterized as a Lorentz violation part and a non-metricity part. For example, in the first case, $B^{\mu\nu} = g^{\mu\nu}+ b^{\mu\nu}$, the covariant derivative of the energy-momentum tensor can be written as 
\begin{align}
    \nabla_\mu T^{\mu}_{\,\,\,\nu}&=\nabla_\mu T^{\mu(LV)}_{\,\,\,\nu} -\frac{(B^{\alpha\beta} \partial_\alpha \phi \partial_\beta \phi)( \nabla_\nu B^{\rho \sigma})\partial_\rho \phi \partial_\sigma \phi}{2M_{LV}},\label{non-conserve}\\
    T^{\mu(LV)}_{\,\,\,\nu} &= -\frac{(B^{\alpha\beta} \partial_\alpha \phi \partial_\beta \phi)( b^{\mu \sigma})\partial_\nu \phi \partial_\sigma \phi}{M_{LV}}
\end{align}
Obviously,  the non-conservative effect becomes relevant in the regime $(\nabla\phi\nabla\phi)>M_{LV}^4$ since $\nabla^\mu T^i_{\mu\nu}=O\left(M_{LV}^{-4}\right)$. The Lorentz-violation part characterized by $T^{(LV)}_{\mu\nu}$ will vanish for the Lorentz-violation tensor, $b^{\mu\nu}$ being zero. The Lorentz-violation feature is not a novel concept. However, this area has long been pioneered in particle physics and cosmology, which can commonly appear in the Rastall gravity \cite{PhysRevD.6.3357} and the LV theories, such as the standard model extension, see \cite{Colladay:1996iz, Colladay:1998fq, PhysRevD.91.065034}. Even the Bran-Dicke theory \cite{Velten:2021xxw}, which can be invariant under the Lorentz symmetry, may exhibit the violation of energy-momentum conservation. In this category, various models \cite{Pretel_2021, Fazlollahi:2023rhg, Fazlollahi:2024lrt} are not entirely excluded from the observational data. Consequently, there is room for further exploration in this direction. For non-metricity part, the second term in the right-hand side of Eq. \eqref{non-conserve}, is characterized by $\nabla_\nu B^{\rho\sigma}$, which vanishes by taking $B^{\rho\sigma} = g^{\mu\nu}$. In this sense, the tensor $b^{\mu\nu}$ also characterizes how the tensor $B^{\mu\nu}$  non-metricitically deviates from the metric tensor. In this direction, there have been intensively investigated the modification of gravitational theory due to non-metricity \cite{BeltranJimenez:2017tkd,BeltranJimenez:2019tme}. These may shed light on the link between an alternative gravitational theory and the ghost condensate aspect.

 \section{Conclusion}
  We found a specific choice of UV complete theory embedded with the Lorentz violation that can stabilize the particle excitation in the phantom field. This modified ghost condensate can provide the particle excitation with positive kinetic energy. The Lorentz invariant of the quadratic term is restored at the low-energy quantum level.  The particle excitation can propagate through the Minkowski spacetime with the traditional relativistic momentum in an arbitrary background value of the phantom field. An unhealthy situation, such as the tree-level gradient instability, can be alleviated.

\section*{Acknowledgments}
This research has received funding support from the NSFR via the Program Management Unit for Human Resource and Institutional Development Research and Innovation [grant number B13F670070]. This research has received funding support from King Mongkut's Institute of Technology Ladkrabang [KREF186713].
 
 \bibliographystyle{apsrev4-1}
\bibliography{mybib.bib}

\begin{thebibliography}{40}%
\makeatletter
\providecommand \@ifxundefined [1]{%
 \@ifx{#1\undefined}
}%
\providecommand \@ifnum [1]{%
 \ifnum #1\expandafter \@firstoftwo
 \else \expandafter \@secondoftwo
 \fi
}%
\providecommand \@ifx [1]{%
 \ifx #1\expandafter \@firstoftwo
 \else \expandafter \@secondoftwo
 \fi
}%
\providecommand \natexlab [1]{#1}%
\providecommand \enquote  [1]{``#1''}%
\providecommand \bibnamefont  [1]{#1}%
\providecommand \bibfnamefont [1]{#1}%
\providecommand \citenamefont [1]{#1}%
\providecommand \href@noop [0]{\@secondoftwo}%
\providecommand \href [0]{\begingroup \@sanitize@url \@href}%
\providecommand \@href[1]{\@@startlink{#1}\@@href}%
\providecommand \@@href[1]{\endgroup#1\@@endlink}%
\providecommand \@sanitize@url [0]{\catcode `\\12\catcode `\$12\catcode `\&12\catcode `\#12\catcode `\^12\catcode `\_12\catcode `\%12\relax}%
\providecommand \@@startlink[1]{}%
\providecommand \@@endlink[0]{}%
\providecommand \url  [0]{\begingroup\@sanitize@url \@url }%
\providecommand \@url [1]{\endgroup\@href {#1}{\urlprefix }}%
\providecommand \urlprefix  [0]{URL }%
\providecommand \Eprint [0]{\href }%
\providecommand \doibase [0]{http://dx.doi.org/}%
\providecommand \selectlanguage [0]{\@gobble}%
\providecommand \bibinfo  [0]{\@secondoftwo}%
\providecommand \bibfield  [0]{\@secondoftwo}%
\providecommand \translation [1]{[#1]}%
\providecommand \BibitemOpen [0]{}%
\providecommand \bibitemStop [0]{}%
\providecommand \bibitemNoStop [0]{.\EOS\space}%
\providecommand \EOS [0]{\spacefactor3000\relax}%
\providecommand \BibitemShut  [1]{\csname bibitem#1\endcsname}%
\let\auto@bib@innerbib\@empty
\bibitem [{\citenamefont {Caldwell}(2002)}]{CALDWELL200223}%
  \BibitemOpen
  \bibfield  {author} {\bibinfo {author} {\bibfnamefont {R.}~\bibnamefont {Caldwell}},\ }\href {\doibase https://doi.org/10.1016/S0370-2693(02)02589-3} {\bibfield  {journal} {\bibinfo  {journal} {Physics Letters B}\ }\textbf {\bibinfo {volume} {545}},\ \bibinfo {pages} {23} (\bibinfo {year} {2002})}\BibitemShut {NoStop}%
\bibitem [{\citenamefont {Ludwick}(2017)}]{Ludwick:2017tox}%
  \BibitemOpen
  \bibfield  {author} {\bibinfo {author} {\bibfnamefont {K.~J.}\ \bibnamefont {Ludwick}},\ }\href {\doibase 10.1142/S0217732317300257} {\bibfield  {journal} {\bibinfo  {journal} {Mod. Phys. Lett. A}\ }\textbf {\bibinfo {volume} {32}},\ \bibinfo {pages} {1730025} (\bibinfo {year} {2017})},\ \Eprint {http://arxiv.org/abs/1708.06981} {arXiv:1708.06981 [astro-ph.CO]} \BibitemShut {NoStop}%
\bibitem [{\citenamefont {Carroll}\ \emph {et~al.}(2003)\citenamefont {Carroll}, \citenamefont {Hoffman},\ and\ \citenamefont {Trodden}}]{PhysRevD.68.023509}%
  \BibitemOpen
  \bibfield  {author} {\bibinfo {author} {\bibfnamefont {S.~M.}\ \bibnamefont {Carroll}}, \bibinfo {author} {\bibfnamefont {M.}~\bibnamefont {Hoffman}}, \ and\ \bibinfo {author} {\bibfnamefont {M.}~\bibnamefont {Trodden}},\ }\href {\doibase 10.1103/PhysRevD.68.023509} {\bibfield  {journal} {\bibinfo  {journal} {Phys. Rev. D}\ }\textbf {\bibinfo {volume} {68}},\ \bibinfo {pages} {023509} (\bibinfo {year} {2003})}\BibitemShut {NoStop}%
\bibitem [{\citenamefont {Vikman}(2005)}]{PhysRevD.71.023515}%
  \BibitemOpen
  \bibfield  {author} {\bibinfo {author} {\bibfnamefont {A.}~\bibnamefont {Vikman}},\ }\href {\doibase 10.1103/PhysRevD.71.023515} {\bibfield  {journal} {\bibinfo  {journal} {Phys. Rev. D}\ }\textbf {\bibinfo {volume} {71}},\ \bibinfo {pages} {023515} (\bibinfo {year} {2005})}\BibitemShut {NoStop}%
\bibitem [{\citenamefont {Nojiri}\ and\ \citenamefont {Odintsov}(2005)}]{PhysRevD.72.023003}%
  \BibitemOpen
  \bibfield  {author} {\bibinfo {author} {\bibfnamefont {S.}~\bibnamefont {Nojiri}}\ and\ \bibinfo {author} {\bibfnamefont {S.~D.}\ \bibnamefont {Odintsov}},\ }\href {\doibase 10.1103/PhysRevD.72.023003} {\bibfield  {journal} {\bibinfo  {journal} {Phys. Rev. D}\ }\textbf {\bibinfo {volume} {72}},\ \bibinfo {pages} {023003} (\bibinfo {year} {2005})}\BibitemShut {NoStop}%
\bibitem [{\citenamefont {Nojiri}\ and\ \citenamefont {Odintsov}(2006{\natexlab{a}})}]{Nojiri:2005pu}%
  \BibitemOpen
  \bibfield  {author} {\bibinfo {author} {\bibfnamefont {S.}~\bibnamefont {Nojiri}}\ and\ \bibinfo {author} {\bibfnamefont {S.~D.}\ \bibnamefont {Odintsov}},\ }\href {\doibase 10.1007/s10714-006-0301-6} {\bibfield  {journal} {\bibinfo  {journal} {Gen. Rel. Grav.}\ }\textbf {\bibinfo {volume} {38}},\ \bibinfo {pages} {1285} (\bibinfo {year} {2006}{\natexlab{a}})},\ \Eprint {http://arxiv.org/abs/hep-th/0506212} {arXiv:hep-th/0506212} \BibitemShut {NoStop}%
\bibitem [{\citenamefont {Nojiri}\ and\ \citenamefont {Odintsov}(2006{\natexlab{b}})}]{Nojiri:2006ww}%
  \BibitemOpen
  \bibfield  {author} {\bibinfo {author} {\bibfnamefont {S.}~\bibnamefont {Nojiri}}\ and\ \bibinfo {author} {\bibfnamefont {S.~D.}\ \bibnamefont {Odintsov}},\ }\href {\doibase 10.1016/j.physletb.2006.04.026} {\bibfield  {journal} {\bibinfo  {journal} {Phys. Lett. B}\ }\textbf {\bibinfo {volume} {637}},\ \bibinfo {pages} {139} (\bibinfo {year} {2006}{\natexlab{b}})},\ \Eprint {http://arxiv.org/abs/hep-th/0603062} {arXiv:hep-th/0603062} \BibitemShut {NoStop}%
\bibitem [{\citenamefont {Di~Valentino}\ \emph {et~al.}(2021)\citenamefont {Di~Valentino}, \citenamefont {Mukherjee},\ and\ \citenamefont {Sen}}]{e23040404}%
  \BibitemOpen
  \bibfield  {author} {\bibinfo {author} {\bibfnamefont {E.}~\bibnamefont {Di~Valentino}}, \bibinfo {author} {\bibfnamefont {A.}~\bibnamefont {Mukherjee}}, \ and\ \bibinfo {author} {\bibfnamefont {A.~A.}\ \bibnamefont {Sen}},\ }\href {\doibase 10.3390/e23040404} {\bibfield  {journal} {\bibinfo  {journal} {Entropy}\ }\textbf {\bibinfo {volume} {23}} (\bibinfo {year} {2021}),\ 10.3390/e23040404}\BibitemShut {NoStop}%
\bibitem [{\citenamefont {Tian}\ and\ \citenamefont {Zhu}(2021)}]{PhysRevD.103.043518}%
  \BibitemOpen
  \bibfield  {author} {\bibinfo {author} {\bibfnamefont {S.~X.}\ \bibnamefont {Tian}}\ and\ \bibinfo {author} {\bibfnamefont {Z.-H.}\ \bibnamefont {Zhu}},\ }\href {\doibase 10.1103/PhysRevD.103.043518} {\bibfield  {journal} {\bibinfo  {journal} {Phys. Rev. D}\ }\textbf {\bibinfo {volume} {103}},\ \bibinfo {pages} {043518} (\bibinfo {year} {2021})}\BibitemShut {NoStop}%
\bibitem [{\citenamefont {Armendariz-Picon}\ \emph {et~al.}(2001)\citenamefont {Armendariz-Picon}, \citenamefont {Mukhanov},\ and\ \citenamefont {Steinhardt}}]{PhysRevD.63.103510}%
  \BibitemOpen
  \bibfield  {author} {\bibinfo {author} {\bibfnamefont {C.}~\bibnamefont {Armendariz-Picon}}, \bibinfo {author} {\bibfnamefont {V.}~\bibnamefont {Mukhanov}}, \ and\ \bibinfo {author} {\bibfnamefont {P.~J.}\ \bibnamefont {Steinhardt}},\ }\href {\doibase 10.1103/PhysRevD.63.103510} {\bibfield  {journal} {\bibinfo  {journal} {Phys. Rev. D}\ }\textbf {\bibinfo {volume} {63}},\ \bibinfo {pages} {103510} (\bibinfo {year} {2001})}\BibitemShut {NoStop}%
\bibitem [{\citenamefont {Armendáriz-Picón}\ \emph {et~al.}(1999)\citenamefont {Armendáriz-Picón}, \citenamefont {Damour},\ and\ \citenamefont {Mukhanov}}]{ARMENDARIZPICON1999209}%
  \BibitemOpen
  \bibfield  {author} {\bibinfo {author} {\bibfnamefont {C.}~\bibnamefont {Armendáriz-Picón}}, \bibinfo {author} {\bibfnamefont {T.}~\bibnamefont {Damour}}, \ and\ \bibinfo {author} {\bibfnamefont {V.}~\bibnamefont {Mukhanov}},\ }\href {\doibase https://doi.org/10.1016/S0370-2693(99)00603-6} {\bibfield  {journal} {\bibinfo  {journal} {Physics Letters B}\ }\textbf {\bibinfo {volume} {458}},\ \bibinfo {pages} {209} (\bibinfo {year} {1999})}\BibitemShut {NoStop}%
\bibitem [{\citenamefont {Cai}\ \emph {et~al.}(2011)\citenamefont {Cai}, \citenamefont {Dent},\ and\ \citenamefont {Easson}}]{PhysRevD.83.101301}%
  \BibitemOpen
  \bibfield  {author} {\bibinfo {author} {\bibfnamefont {Y.-F.}\ \bibnamefont {Cai}}, \bibinfo {author} {\bibfnamefont {J.~B.}\ \bibnamefont {Dent}}, \ and\ \bibinfo {author} {\bibfnamefont {D.~A.}\ \bibnamefont {Easson}},\ }\href {\doibase 10.1103/PhysRevD.83.101301} {\bibfield  {journal} {\bibinfo  {journal} {Phys. Rev. D}\ }\textbf {\bibinfo {volume} {83}},\ \bibinfo {pages} {101301} (\bibinfo {year} {2011})}\BibitemShut {NoStop}%
\bibitem [{\citenamefont {Alishahiha}\ \emph {et~al.}(2004)\citenamefont {Alishahiha}, \citenamefont {Silverstein},\ and\ \citenamefont {Tong}}]{PhysRevD.70.123505}%
  \BibitemOpen
  \bibfield  {author} {\bibinfo {author} {\bibfnamefont {M.}~\bibnamefont {Alishahiha}}, \bibinfo {author} {\bibfnamefont {E.}~\bibnamefont {Silverstein}}, \ and\ \bibinfo {author} {\bibfnamefont {D.}~\bibnamefont {Tong}},\ }\href {\doibase 10.1103/PhysRevD.70.123505} {\bibfield  {journal} {\bibinfo  {journal} {Phys. Rev. D}\ }\textbf {\bibinfo {volume} {70}},\ \bibinfo {pages} {123505} (\bibinfo {year} {2004})}\BibitemShut {NoStop}%
\bibitem [{\citenamefont {Huang}\ \emph {et~al.}(2008)\citenamefont {Huang}, \citenamefont {Shiu},\ and\ \citenamefont {Underwood}}]{PhysRevD.77.023511}%
  \BibitemOpen
  \bibfield  {author} {\bibinfo {author} {\bibfnamefont {M.-x.}\ \bibnamefont {Huang}}, \bibinfo {author} {\bibfnamefont {G.}~\bibnamefont {Shiu}}, \ and\ \bibinfo {author} {\bibfnamefont {B.}~\bibnamefont {Underwood}},\ }\href {\doibase 10.1103/PhysRevD.77.023511} {\bibfield  {journal} {\bibinfo  {journal} {Phys. Rev. D}\ }\textbf {\bibinfo {volume} {77}},\ \bibinfo {pages} {023511} (\bibinfo {year} {2008})}\BibitemShut {NoStop}%
\bibitem [{\citenamefont {Hu}(2005)}]{PhysRevD.71.047301}%
  \BibitemOpen
  \bibfield  {author} {\bibinfo {author} {\bibfnamefont {W.}~\bibnamefont {Hu}},\ }\href {\doibase 10.1103/PhysRevD.71.047301} {\bibfield  {journal} {\bibinfo  {journal} {Phys. Rev. D}\ }\textbf {\bibinfo {volume} {71}},\ \bibinfo {pages} {047301} (\bibinfo {year} {2005})}\BibitemShut {NoStop}%
\bibitem [{\citenamefont {Zhao}\ and\ \citenamefont {Zhang}(2006)}]{PhysRevD.73.123509}%
  \BibitemOpen
  \bibfield  {author} {\bibinfo {author} {\bibfnamefont {W.}~\bibnamefont {Zhao}}\ and\ \bibinfo {author} {\bibfnamefont {Y.}~\bibnamefont {Zhang}},\ }\href {\doibase 10.1103/PhysRevD.73.123509} {\bibfield  {journal} {\bibinfo  {journal} {Phys. Rev. D}\ }\textbf {\bibinfo {volume} {73}},\ \bibinfo {pages} {123509} (\bibinfo {year} {2006})}\BibitemShut {NoStop}%
\bibitem [{\citenamefont {Cai}\ \emph {et~al.}(2010)\citenamefont {Cai}, \citenamefont {Saridakis}, \citenamefont {Setare},\ and\ \citenamefont {Xia}}]{Cai:2009zp}%
  \BibitemOpen
  \bibfield  {author} {\bibinfo {author} {\bibfnamefont {Y.-F.}\ \bibnamefont {Cai}}, \bibinfo {author} {\bibfnamefont {E.~N.}\ \bibnamefont {Saridakis}}, \bibinfo {author} {\bibfnamefont {M.~R.}\ \bibnamefont {Setare}}, \ and\ \bibinfo {author} {\bibfnamefont {J.-Q.}\ \bibnamefont {Xia}},\ }\href {\doibase 10.1016/j.physrep.2010.04.001} {\bibfield  {journal} {\bibinfo  {journal} {Phys. Rept.}\ }\textbf {\bibinfo {volume} {493}},\ \bibinfo {pages} {1} (\bibinfo {year} {2010})},\ \Eprint {http://arxiv.org/abs/0909.2776} {arXiv:0909.2776 [hep-th]} \BibitemShut {NoStop}%
\bibitem [{\citenamefont {Yang}\ \emph {et~al.}(2024)\citenamefont {Yang}, \citenamefont {Ren}, \citenamefont {Wang}, \citenamefont {Lu}, \citenamefont {Zhang}, \citenamefont {Cai},\ and\ \citenamefont {Saridakis}}]{Yang:2024kdo}%
  \BibitemOpen
  \bibfield  {author} {\bibinfo {author} {\bibfnamefont {Y.}~\bibnamefont {Yang}}, \bibinfo {author} {\bibfnamefont {X.}~\bibnamefont {Ren}}, \bibinfo {author} {\bibfnamefont {Q.}~\bibnamefont {Wang}}, \bibinfo {author} {\bibfnamefont {Z.}~\bibnamefont {Lu}}, \bibinfo {author} {\bibfnamefont {D.}~\bibnamefont {Zhang}}, \bibinfo {author} {\bibfnamefont {Y.-F.}\ \bibnamefont {Cai}}, \ and\ \bibinfo {author} {\bibfnamefont {E.~N.}\ \bibnamefont {Saridakis}},\ }\href {\doibase 10.1016/j.scib.2024.07.029} {\bibfield  {journal} {\bibinfo  {journal} {Sci. Bull.}\ }\textbf {\bibinfo {volume} {69}},\ \bibinfo {pages} {2698} (\bibinfo {year} {2024})},\ \Eprint {http://arxiv.org/abs/2404.19437} {arXiv:2404.19437 [astro-ph.CO]} \BibitemShut {NoStop}%
\bibitem [{\citenamefont {Arkani-Hamed}\ \emph {et~al.}(2004{\natexlab{a}})\citenamefont {Arkani-Hamed}, \citenamefont {Cheng}, \citenamefont {Luty},\ and\ \citenamefont {Mukohyama}}]{Arkani-Hamed:2003pdi}%
  \BibitemOpen
  \bibfield  {author} {\bibinfo {author} {\bibfnamefont {N.}~\bibnamefont {Arkani-Hamed}}, \bibinfo {author} {\bibfnamefont {H.-C.}\ \bibnamefont {Cheng}}, \bibinfo {author} {\bibfnamefont {M.~A.}\ \bibnamefont {Luty}}, \ and\ \bibinfo {author} {\bibfnamefont {S.}~\bibnamefont {Mukohyama}},\ }\href {\doibase 10.1088/1126-6708/2004/05/074} {\bibfield  {journal} {\bibinfo  {journal} {JHEP}\ }\textbf {\bibinfo {volume} {05}},\ \bibinfo {pages} {074} (\bibinfo {year} {2004}{\natexlab{a}})},\ \Eprint {http://arxiv.org/abs/hep-th/0312099} {arXiv:hep-th/0312099} \BibitemShut {NoStop}%
\bibitem [{\citenamefont {Arkani-Hamed}\ \emph {et~al.}(2004{\natexlab{b}})\citenamefont {Arkani-Hamed}, \citenamefont {Creminelli}, \citenamefont {Mukohyama},\ and\ \citenamefont {Zaldarriaga}}]{Arkani-Hamed:2003juy}%
  \BibitemOpen
  \bibfield  {author} {\bibinfo {author} {\bibfnamefont {N.}~\bibnamefont {Arkani-Hamed}}, \bibinfo {author} {\bibfnamefont {P.}~\bibnamefont {Creminelli}}, \bibinfo {author} {\bibfnamefont {S.}~\bibnamefont {Mukohyama}}, \ and\ \bibinfo {author} {\bibfnamefont {M.}~\bibnamefont {Zaldarriaga}},\ }\href {\doibase 10.1088/1475-7516/2004/04/001} {\bibfield  {journal} {\bibinfo  {journal} {JCAP}\ }\textbf {\bibinfo {volume} {04}},\ \bibinfo {pages} {001} (\bibinfo {year} {2004}{\natexlab{b}})},\ \Eprint {http://arxiv.org/abs/hep-th/0312100} {arXiv:hep-th/0312100} \BibitemShut {NoStop}%
\bibitem [{\citenamefont {Anisimov}\ and\ \citenamefont {Vikman}(2005)}]{Anisimov:2004sp}%
  \BibitemOpen
  \bibfield  {author} {\bibinfo {author} {\bibfnamefont {A.}~\bibnamefont {Anisimov}}\ and\ \bibinfo {author} {\bibfnamefont {A.}~\bibnamefont {Vikman}},\ }\href {\doibase 10.1088/1475-7516/2005/04/009} {\bibfield  {journal} {\bibinfo  {journal} {JCAP}\ }\textbf {\bibinfo {volume} {04}},\ \bibinfo {pages} {009} (\bibinfo {year} {2005})},\ \Eprint {http://arxiv.org/abs/hep-ph/0411089} {arXiv:hep-ph/0411089} \BibitemShut {NoStop}%
\bibitem [{\citenamefont {Piazza}\ and\ \citenamefont {Tsujikawa}(2004)}]{Piazza:2004df}%
  \BibitemOpen
  \bibfield  {author} {\bibinfo {author} {\bibfnamefont {F.}~\bibnamefont {Piazza}}\ and\ \bibinfo {author} {\bibfnamefont {S.}~\bibnamefont {Tsujikawa}},\ }\href {\doibase 10.1088/1475-7516/2004/07/004} {\bibfield  {journal} {\bibinfo  {journal} {JCAP}\ }\textbf {\bibinfo {volume} {07}},\ \bibinfo {pages} {004} (\bibinfo {year} {2004})},\ \Eprint {http://arxiv.org/abs/hep-th/0405054} {arXiv:hep-th/0405054} \BibitemShut {NoStop}%
\bibitem [{\citenamefont {Arkani-Hamed}\ \emph {et~al.}(2007)\citenamefont {Arkani-Hamed}, \citenamefont {Cheng}, \citenamefont {Luty}, \citenamefont {Mukohyama},\ and\ \citenamefont {Wiseman}}]{Arkani-Hamed:2005teg}%
  \BibitemOpen
  \bibfield  {author} {\bibinfo {author} {\bibfnamefont {N.}~\bibnamefont {Arkani-Hamed}}, \bibinfo {author} {\bibfnamefont {H.-C.}\ \bibnamefont {Cheng}}, \bibinfo {author} {\bibfnamefont {M.~A.}\ \bibnamefont {Luty}}, \bibinfo {author} {\bibfnamefont {S.}~\bibnamefont {Mukohyama}}, \ and\ \bibinfo {author} {\bibfnamefont {T.}~\bibnamefont {Wiseman}},\ }\href {\doibase 10.1088/1126-6708/2007/01/036} {\bibfield  {journal} {\bibinfo  {journal} {JHEP}\ }\textbf {\bibinfo {volume} {01}},\ \bibinfo {pages} {036} (\bibinfo {year} {2007})},\ \Eprint {http://arxiv.org/abs/hep-ph/0507120} {arXiv:hep-ph/0507120} \BibitemShut {NoStop}%
\bibitem [{\citenamefont {Bhattacharya}\ \emph {et~al.}(2015)\citenamefont {Bhattacharya}, \citenamefont {Mukherjee}, \citenamefont {Singha~Roy},\ and\ \citenamefont {Saha}}]{Bhattacharya:2014pta}%
  \BibitemOpen
  \bibfield  {author} {\bibinfo {author} {\bibfnamefont {G.}~\bibnamefont {Bhattacharya}}, \bibinfo {author} {\bibfnamefont {P.}~\bibnamefont {Mukherjee}}, \bibinfo {author} {\bibfnamefont {A.}~\bibnamefont {Singha~Roy}}, \ and\ \bibinfo {author} {\bibfnamefont {A.}~\bibnamefont {Saha}},\ }\href {\doibase 10.1140/epjc/s10052-015-3301-4} {\bibfield  {journal} {\bibinfo  {journal} {Eur. Phys. J. C}\ }\textbf {\bibinfo {volume} {75}},\ \bibinfo {pages} {84} (\bibinfo {year} {2015})},\ \Eprint {http://arxiv.org/abs/1401.6745} {arXiv:1401.6745 [gr-qc]} \BibitemShut {NoStop}%
\bibitem [{\citenamefont {Colladay}\ and\ \citenamefont {Kostelecky}(1997)}]{Colladay:1996iz}%
  \BibitemOpen
  \bibfield  {author} {\bibinfo {author} {\bibfnamefont {D.}~\bibnamefont {Colladay}}\ and\ \bibinfo {author} {\bibfnamefont {V.~A.}\ \bibnamefont {Kostelecky}},\ }\href {\doibase 10.1103/PhysRevD.55.6760} {\bibfield  {journal} {\bibinfo  {journal} {Phys. Rev. D}\ }\textbf {\bibinfo {volume} {55}},\ \bibinfo {pages} {6760} (\bibinfo {year} {1997})},\ \Eprint {http://arxiv.org/abs/hep-ph/9703464} {arXiv:hep-ph/9703464} \BibitemShut {NoStop}%
\bibitem [{\citenamefont {Colladay}\ and\ \citenamefont {Kostelecky}(1998)}]{Colladay:1998fq}%
  \BibitemOpen
  \bibfield  {author} {\bibinfo {author} {\bibfnamefont {D.}~\bibnamefont {Colladay}}\ and\ \bibinfo {author} {\bibfnamefont {V.~A.}\ \bibnamefont {Kostelecky}},\ }\href {\doibase 10.1103/PhysRevD.58.116002} {\bibfield  {journal} {\bibinfo  {journal} {Phys. Rev. D}\ }\textbf {\bibinfo {volume} {58}},\ \bibinfo {pages} {116002} (\bibinfo {year} {1998})},\ \Eprint {http://arxiv.org/abs/hep-ph/9809521} {arXiv:hep-ph/9809521} \BibitemShut {NoStop}%
\bibitem [{\citenamefont {Abazov}\ \emph {et~al.}(2012)\citenamefont {Abazov} \emph {et~al.}}]{D0:2012rbu}%
  \BibitemOpen
  \bibfield  {author} {\bibinfo {author} {\bibfnamefont {V.~M.}\ \bibnamefont {Abazov}} \emph {et~al.} (\bibinfo {collaboration} {D0}),\ }\href {\doibase 10.1103/PhysRevLett.108.261603} {\bibfield  {journal} {\bibinfo  {journal} {Phys. Rev. Lett.}\ }\textbf {\bibinfo {volume} {108}},\ \bibinfo {pages} {261603} (\bibinfo {year} {2012})},\ \Eprint {http://arxiv.org/abs/1203.6106} {arXiv:1203.6106 [hep-ex]} \BibitemShut {NoStop}%
\bibitem [{\citenamefont {Carle}\ \emph {et~al.}(2020)\citenamefont {Carle}, \citenamefont {Chanon},\ and\ \citenamefont {Perries}}]{Carle:2019ouy}%
  \BibitemOpen
  \bibfield  {author} {\bibinfo {author} {\bibfnamefont {A.}~\bibnamefont {Carle}}, \bibinfo {author} {\bibfnamefont {N.}~\bibnamefont {Chanon}}, \ and\ \bibinfo {author} {\bibfnamefont {S.}~\bibnamefont {Perries}},\ }\href {\doibase 10.1140/epjc/s10052-020-7715-2} {\bibfield  {journal} {\bibinfo  {journal} {Eur. Phys. J. C}\ }\textbf {\bibinfo {volume} {80}},\ \bibinfo {pages} {128} (\bibinfo {year} {2020})},\ \Eprint {http://arxiv.org/abs/1908.11256} {arXiv:1908.11256 [hep-ph]} \BibitemShut {NoStop}%
\bibitem [{\citenamefont {Hayrapetyan}\ \emph {et~al.}(2024)\citenamefont {Hayrapetyan} \emph {et~al.}}]{CMS:2024rcv}%
  \BibitemOpen
  \bibfield  {author} {\bibinfo {author} {\bibfnamefont {A.}~\bibnamefont {Hayrapetyan}} \emph {et~al.} (\bibinfo {collaboration} {CMS}),\ }\href {\doibase 10.1016/j.physletb.2024.138979} {\bibfield  {journal} {\bibinfo  {journal} {Phys. Lett. B}\ }\textbf {\bibinfo {volume} {857}},\ \bibinfo {pages} {138979} (\bibinfo {year} {2024})},\ \Eprint {http://arxiv.org/abs/2405.14757} {arXiv:2405.14757 [hep-ex]} \BibitemShut {NoStop}%
\bibitem [{\citenamefont {Hsu}\ \emph {et~al.}(2004)\citenamefont {Hsu}, \citenamefont {Jenkins},\ and\ \citenamefont {Wise}}]{Hsu:2004vr}%
  \BibitemOpen
  \bibfield  {author} {\bibinfo {author} {\bibfnamefont {S.~D.~H.}\ \bibnamefont {Hsu}}, \bibinfo {author} {\bibfnamefont {A.}~\bibnamefont {Jenkins}}, \ and\ \bibinfo {author} {\bibfnamefont {M.~B.}\ \bibnamefont {Wise}},\ }\href {\doibase 10.1016/j.physletb.2004.07.025} {\bibfield  {journal} {\bibinfo  {journal} {Phys. Lett. B}\ }\textbf {\bibinfo {volume} {597}},\ \bibinfo {pages} {270} (\bibinfo {year} {2004})},\ \Eprint {http://arxiv.org/abs/astro-ph/0406043} {arXiv:astro-ph/0406043} \BibitemShut {NoStop}%
\bibitem [{\citenamefont {Manohar}\ and\ \citenamefont {Georgi}(1984)}]{Manohar:1983md}%
  \BibitemOpen
  \bibfield  {author} {\bibinfo {author} {\bibfnamefont {A.}~\bibnamefont {Manohar}}\ and\ \bibinfo {author} {\bibfnamefont {H.}~\bibnamefont {Georgi}},\ }\href {\doibase 10.1016/0550-3213(84)90231-1} {\bibfield  {journal} {\bibinfo  {journal} {Nucl. Phys. B}\ }\textbf {\bibinfo {volume} {234}},\ \bibinfo {pages} {189} (\bibinfo {year} {1984})}\BibitemShut {NoStop}%
\bibitem [{\citenamefont {Gavela}\ \emph {et~al.}(2016)\citenamefont {Gavela}, \citenamefont {Jenkins}, \citenamefont {Manohar},\ and\ \citenamefont {Merlo}}]{Gavela:2016bzc}%
  \BibitemOpen
  \bibfield  {author} {\bibinfo {author} {\bibfnamefont {B.~M.}\ \bibnamefont {Gavela}}, \bibinfo {author} {\bibfnamefont {E.~E.}\ \bibnamefont {Jenkins}}, \bibinfo {author} {\bibfnamefont {A.~V.}\ \bibnamefont {Manohar}}, \ and\ \bibinfo {author} {\bibfnamefont {L.}~\bibnamefont {Merlo}},\ }\href {\doibase 10.1140/epjc/s10052-016-4332-1} {\bibfield  {journal} {\bibinfo  {journal} {Eur. Phys. J. C}\ }\textbf {\bibinfo {volume} {76}},\ \bibinfo {pages} {485} (\bibinfo {year} {2016})},\ \Eprint {http://arxiv.org/abs/1601.07551} {arXiv:1601.07551 [hep-ph]} \BibitemShut {NoStop}%
\bibitem [{\citenamefont {Rastall}(1972)}]{PhysRevD.6.3357}%
  \BibitemOpen
  \bibfield  {author} {\bibinfo {author} {\bibfnamefont {P.}~\bibnamefont {Rastall}},\ }\href {\doibase 10.1103/PhysRevD.6.3357} {\bibfield  {journal} {\bibinfo  {journal} {Phys. Rev. D}\ }\textbf {\bibinfo {volume} {6}},\ \bibinfo {pages} {3357} (\bibinfo {year} {1972})}\BibitemShut {NoStop}%
\bibitem [{\citenamefont {Bluhm}(2015)}]{PhysRevD.91.065034}%
  \BibitemOpen
  \bibfield  {author} {\bibinfo {author} {\bibfnamefont {R.}~\bibnamefont {Bluhm}},\ }\href {\doibase 10.1103/PhysRevD.91.065034} {\bibfield  {journal} {\bibinfo  {journal} {Phys. Rev. D}\ }\textbf {\bibinfo {volume} {91}},\ \bibinfo {pages} {065034} (\bibinfo {year} {2015})}\BibitemShut {NoStop}%
\bibitem [{\citenamefont {Velten}\ and\ \citenamefont {Caram\^es}(2021)}]{Velten:2021xxw}%
  \BibitemOpen
  \bibfield  {author} {\bibinfo {author} {\bibfnamefont {H.}~\bibnamefont {Velten}}\ and\ \bibinfo {author} {\bibfnamefont {T.~R.~P.}\ \bibnamefont {Caram\^es}},\ }\href {\doibase 10.3390/universe7020038} {\bibfield  {journal} {\bibinfo  {journal} {Universe}\ }\textbf {\bibinfo {volume} {7}},\ \bibinfo {pages} {38} (\bibinfo {year} {2021})},\ \Eprint {http://arxiv.org/abs/2102.03457} {arXiv:2102.03457 [gr-qc]} \BibitemShut {NoStop}%
\bibitem [{\citenamefont {Pretel}\ \emph {et~al.}(2021)\citenamefont {Pretel}, \citenamefont {Jorás}, \citenamefont {Reis},\ and\ \citenamefont {Arbañil}}]{Pretel_2021}%
  \BibitemOpen
  \bibfield  {author} {\bibinfo {author} {\bibfnamefont {J.~M.}\ \bibnamefont {Pretel}}, \bibinfo {author} {\bibfnamefont {S.~E.}\ \bibnamefont {Jorás}}, \bibinfo {author} {\bibfnamefont {R.~R.}\ \bibnamefont {Reis}}, \ and\ \bibinfo {author} {\bibfnamefont {J.~D.}\ \bibnamefont {Arbañil}},\ }\href {\doibase 10.1088/1475-7516/2021/08/055} {\bibfield  {journal} {\bibinfo  {journal} {Journal of Cosmology and Astroparticle Physics}\ }\textbf {\bibinfo {volume} {2021}},\ \bibinfo {pages} {055} (\bibinfo {year} {2021})}\BibitemShut {NoStop}%
\bibitem [{\citenamefont {Fazlollahi}(2023)}]{Fazlollahi:2023rhg}%
  \BibitemOpen
  \bibfield  {author} {\bibinfo {author} {\bibfnamefont {H.~R.}\ \bibnamefont {Fazlollahi}},\ }\href {\doibase 10.1140/epjc/s10052-023-12003-x} {\bibfield  {journal} {\bibinfo  {journal} {Eur. Phys. J. C}\ }\textbf {\bibinfo {volume} {83}},\ \bibinfo {pages} {923} (\bibinfo {year} {2023})}\BibitemShut {NoStop}%
\bibitem [{\citenamefont {Fazlollahi}(2024)}]{Fazlollahi:2024lrt}%
  \BibitemOpen
  \bibfield  {author} {\bibinfo {author} {\bibfnamefont {H.~R.}\ \bibnamefont {Fazlollahi}},\ }\href {\doibase 10.1088/1572-9494/ad30f6} {\bibfield  {journal} {\bibinfo  {journal} {Commun. Theor. Phys.}\ }\textbf {\bibinfo {volume} {76}},\ \bibinfo {pages} {045402} (\bibinfo {year} {2024})}\BibitemShut {NoStop}%
\bibitem [{\citenamefont {Beltr{\'a}n~Jim{\'e}nez}\ \emph {et~al.}(2018)\citenamefont {Beltr{\'a}n~Jim{\'e}nez}, \citenamefont {Heisenberg},\ and\ \citenamefont {Koivisto}}]{BeltranJimenez:2017tkd}%
  \BibitemOpen
  \bibfield  {author} {\bibinfo {author} {\bibfnamefont {J.}~\bibnamefont {Beltr{\'a}n~Jim{\'e}nez}}, \bibinfo {author} {\bibfnamefont {L.}~\bibnamefont {Heisenberg}}, \ and\ \bibinfo {author} {\bibfnamefont {T.}~\bibnamefont {Koivisto}},\ }\href {\doibase 10.1103/PhysRevD.98.044048} {\bibfield  {journal} {\bibinfo  {journal} {Phys. Rev. D}\ }\textbf {\bibinfo {volume} {98}},\ \bibinfo {pages} {044048} (\bibinfo {year} {2018})},\ \Eprint {http://arxiv.org/abs/1710.03116} {arXiv:1710.03116 [gr-qc]} \BibitemShut {NoStop}%
\bibitem [{\citenamefont {Beltr{\'a}n~Jim{\'e}nez}\ \emph {et~al.}(2020)\citenamefont {Beltr{\'a}n~Jim{\'e}nez}, \citenamefont {Heisenberg}, \citenamefont {Koivisto},\ and\ \citenamefont {Pekar}}]{BeltranJimenez:2019tme}%
  \BibitemOpen
  \bibfield  {author} {\bibinfo {author} {\bibfnamefont {J.}~\bibnamefont {Beltr{\'a}n~Jim{\'e}nez}}, \bibinfo {author} {\bibfnamefont {L.}~\bibnamefont {Heisenberg}}, \bibinfo {author} {\bibfnamefont {T.~S.}\ \bibnamefont {Koivisto}}, \ and\ \bibinfo {author} {\bibfnamefont {S.}~\bibnamefont {Pekar}},\ }\href {\doibase 10.1103/PhysRevD.101.103507} {\bibfield  {journal} {\bibinfo  {journal} {Phys. Rev. D}\ }\textbf {\bibinfo {volume} {101}},\ \bibinfo {pages} {103507} (\bibinfo {year} {2020})},\ \Eprint {http://arxiv.org/abs/1906.10027} {arXiv:1906.10027 [gr-qc]} \BibitemShut {NoStop}%
\end{thebibliography}%
\end{document}